\documentclass[journal]{IEEEtran}
\usepackage{amsmath}
\usepackage{amsfonts,amssymb}
\usepackage{cite}
\usepackage{bbm}
\usepackage{color,soul}
\usepackage{gensymb}
\usepackage{dsfont}
\usepackage{amsthm}
\usepackage{bm}
\usepackage{algorithm}
\usepackage{algorithmic}

\usepackage{pgfplots}
\usepackage[utf8]{inputenc}
\usepackage{comment}

\ifCLASSOPTIONcompsoc
  \usepackage[caption=false,font=normalsize,labelfont=sf,textfont=sf]{subfig}
\else
  \usepackage[caption=false,font=footnotesize]{subfig}
\fi
\usepackage{euscript}
\allowdisplaybreaks[4]

\ifCLASSOPTIONcompsoc
  \usepackage[caption=false,font=normalsize,labelfont=sf,textfont=sf]{subfig}
\else
  \usepackage[caption=false,font=footnotesize]{subfig}
\fi

\usepackage{url}

\begin{document}
\title{Distribution Grid Modeling \\Using Smart Meter Data }
\author{
Yifei~Guo,~\IEEEmembership{Member,~IEEE,}
Yuxuan Yuan,~\IEEEmembership{Graduate Student Member,~IEEE,}
Zhaoyu~Wang,~\IEEEmembership{Senior Member,~IEEE}
\thanks{This work was supported in part by the U.S. Department of Energy Office of Electricity under DE-OE0000875, and in part by the National Science Foundation under ECCS 1929975. (\textit{Corresponding author: Zhaoyu Wang})

The authors are with the Department of Electrical and Computer Engineering, Iowa State University, Ames, IA 50011, USA (email: yfguo.sdu@gmail.com; yuanyx@iastate.edu; wzy@iastate.edu). }\vspace{-5mm}
}

\markboth{Submitted to IEEE for possible publication. Copyright may be transferred without notice}%
{Shell \MakeLowercase{\textit{et al.}}: Bare Demo of IEEEtran.cls for Journals}

\maketitle

\begin{abstract}
The knowledge of distribution grid models, including topologies and line impedances, is essential to grid monitoring, control and protection. However, this information is often unavailable, incomplete or outdated. The increasing deployment of smart meters (SMs) provides a unique opportunity to address this issue. % However, SMs can only provide low-resolution measurements without phase angle information; and the variability and uncertainty of distributed energy resources (DERs) further complicate the grid modeling.
This paper proposes a two-stage data-driven framework for distribution grid modeling using SM data. In the first stage, we propose to identify the topology via reconstructing a weighted Laplacian matrix of distribution networks, which is mathematically proven to be robust against moderately heterogeneous R/X profiles. 
In the second stage, we develop nonlinear least absolute deviations (LAD) and least squares (LS) regression models to estimate line impedances of single branches based on a nonlinear inverse power flow, which is then embedded within a bottom-up sweep algorithm to achieve the identification across the network in a branch-wise manner. Because the estimation models are inherently non-convex programs and NP-hard, we specially address their tractable convex relaxations and verify the exactness. In addition, we design a conductor library to significantly narrow down the solution space. Numerical results on the modified IEEE 13-bus, 37-bus and 69-bus test feeders validate the effectiveness of the proposed methods.
%that consists of two models: robust topology identification and line parameter estimation, using only smart meter (SM) data. The uniqueness of our method is its computational efficiency and robustness to heterogeneous $R/X$ ratios and measurement errors. Specifically, in the topology identification model, the connectivity of different system nodes is obtained by distinguishing zero and negative non-diagonal entries of the weighted Lapalacian matrix rather than accurately estimating each entry. Then, in the parameter estimation model, a line conductor and construction library is utilized as a constraint to narrow down the search space. Meanwhile, our method uses the full nonlinear model to represent branch flow and designs a bottom-up alternating solution to enable parallel computation, thus further reducing the computation burden. The algorithms are validated in numerical experiments on a number of test cases and real utility data.
\end{abstract}

% Note that keywords are not normally used for perreview papers.

\section{Introduction}\label{introduction}
With an increasing penetration of distributed energy resources (DERs), grid monitoring and energy management is imperative to distribution system operation \cite{kaveh2018}. However, such functionalities require complete and accurate knowledge of distribution grid models, including network topologies and line parameters. Unlike transmission systems that enjoy a high level of data redundancy, distribution grid models could be inaccurate or even unavailable \cite{YW2019}. Some utilities only have simple one-line diagrams of their systems without any detailed line parameters; other utilities may have system models, but they are often incomplete or outdated due to the frequent system expansion and reconfiguration. Field inspection is a conventional approach to draw the model information, which is laborious, costly, and time-consuming, especially for large-scale systems \cite{FCL2014}. This suggests an urgent need of efficient and tractable approaches for distribution grid modeling.

In recent years, the deployment of advanced monitoring and metering infrastructures, e.g., smart meters (SMs) and micro-phasor measurement units ($\mu$PMUs), provides an opportunity to extract the distribution grid models from field measurements \cite{eiasmart}. Some studies have made efforts to explore data-driven approaches for network topology and/or parameter identification. These methods can be roughly classified into two categories according to whether they require (voltage and/or current) phase angle information.

The studies of the first category rely on high-granularity synchrophasor measurements \cite{BD2011,lc2020,JY2018,oa2019}. In \cite{BD2011}, a multi-run optimization method was proposed to estimate line parameters of a three-phase distribution feeder based on the synchronized voltage phasors and line flow measurements. The authors in \cite{lc2020} proposed to identify network topology based on both fundamental and harmonic synchrophasor data, by solving a mixed-integer linear program. With the help of phase angle information, the work of \cite{JY2018} jointly estimated the network topology and parameters by directly reconstructing the admittance matrix. In \cite{oa2019}, a similar joint estimation was achieved by carrying out the topology and parameter identification alternatively. Note that these phasor-based methods required a high or even full coverage of $\mu$PMUs, which is cost-prohibitive, especially for low-voltage grids. In addition, the existing joint topology and parameter estimation methods were developed in a system-wide fashion, where the computational complexity grows significantly with large network sizes.
%In \cite{oa2019}, a least-squares (LS) regression model with sparsity-based regularization was built to recover the admittance matrix of distribution circuits, resorting to a full coverage of $\mu$PMUs. The work \cite{JY2018} exploited multiple data sources, including $\mu$PMUs and inverter sensors, to jointly estimate grid topology and line parameters, wherein a generalized low-rank approximation problem was formulated. One critical shortcoming of Class I models is that they build on the availability of data assumptions, such as high-granularity voltage and/or current phasor measurements from $\mu$PMUs. However, a full coverage of such granular sensors is cost-prohibitive, especially in low-voltage grids. In addition, most of them are developed in a system-wide fashion, of which the computational complexity significantly grows with network size.

Rather than using synchrophasor data, another line of research managed to identify topology or parameters using voltage magnitude and power measurements \cite{ZT2016,dd2018,WL2015,SB2013,WW2020,AM2016}. In \cite{ZT2016}, a mixed-integer quadratic programming (QP) model was developed to identify network topology with the known line impedance information. In \cite{dd2018}, a structure learning method was developed to estimate the grid topology by assuming the nodal power injections were uncorrelated or with non-negative covariances. In \cite{WL2015}, a correlation-based algorithm was proposed to identify the grid topology using SM data, under the assumption that customers' voltage profiles were perfectly correlated because of their short electrical distance. In \cite{SB2013}, a Markov random field-based algorithm was proposed to detect the topology based on uncorrelated power loads. Notice that such statistical assumptions may be challenged by the high penetration of behind-the-meter DERs. The authors in \cite{WW2020} formulated the parameter identification problem as a maximum likelihood estimation model based on the linearized power flow. In \cite{AM2016}, an error compensation model was developed to achieve a robust estimation of distribution line parameters. It is observed that the existing methods in this category either conduct topology identification by using some prior line parameter information (e.g., impedance or R/X ratio), or perform parameter identification with the known topology. A joint network topology and parameter estimation is still challenging in the sense that there is not much prior information available in practice. Furthermore, the linearized power flow model is often used, which cannot guarantee the accuracy, especially when the problem is ill-posed \cite{JY2017}. 

These issues inspire a novel two-stage framework to identify network topology and parameters using SM data without making any assumptions on statistical correlations or prior information on topology and R/X ratios. In the first stage, we develop a robust topology identification method, which consists of a linear least squares (LS) model for estimating a weighted Laplacian matrix and a density-based clustering method for topology recovery. In the second stage, the nonlinear least absolute deviations (LAD) and the nonlinear LS regression models are proposed for parameter estimation of a single branch based on the full nonlinear power flow law and a conductor library that is usually available in practice. Then, a bottom-up sweep algorithm is designed to achieve the parameter identification across the entire system, which carries out the computation of line parameters and line flows, alternatively.
Compared to existing studies, the paper has the following contributions:
\begin{itemize}
\item The proposed topology identification method is initially established on homogeneous distribution grids and is then rigorously proven to be robust against moderately heterogeneous R/X ratios. 
\item To improve the accuracy of parameter estimation, we develop the nonlinear (non-convex) regression models based on the exact inverse power flow, and then explore their convexifications; in addition, a conductor library is exploited to reduce the solution space. These new features render the estimation models effective and tractable.
\item The parameter estimation method enjoys good scalability, in the sense that it is performed in a branch-wise manner, so that the computational complexity is \emph{linear} with the network size (number of nodes/branches).\vspace{0mm}
\end{itemize}
%Note that the proposed method requires the fully observable distribution systems (having measurements for all nodes). For partially observable distribution systems, a pseudo-measurement generation method can be applied \cite{yuanyx}, which is out-of-scope of this paper. It should be noted that a similar strategy has been utilized in previous state estimation-based topology identification works. 

%The rest of this paper is organized as follows. Section II gives the power flow model and introduces the inverse power flow. Sections III and IV present the topology identification and line impedance estimation methods, respectively. Numerical results are given in Section V, followed by conclusions. Some necessary derivations and proofs are provided in Appendix.
\emph{Notations:} For a column vector, let $\|\ast\|_1$ and $\|\ast\|_2$  be its $\mathcal{L}_1$-norm and $\mathcal{L}_2$-norm, respectively. For a matrix, let $[\ast]_i$ and  $[\ast]^j$ denote its $i$-th row and $j$-th column; $[\ast]^{-1}$, $[\ast]^{T}$ and $[\ast]^{-T}$ denote its inverse, transpose and inverse transpose, respectively. Let ${\bf 1}_n$ be the $n\times 1$ column vector with all entries being 1 and ${\bf I}_n$ be the $n\times n$ identity matrix.
\section{Preliminaries}
Consider a radial distribution network comprised of $n+1$ buses. Let $\mathcal{N}\bigcup\{0\}$ be the set of bus where the secondary bus of substation transformer is indexed by 0 and $\mathcal{N}:=\{1,...,n\}$ denotes other buses. For $i\in\mathcal{N}$, $\mathcal{C}_i\subseteq\mathcal{N}$ denotes its children bus set. $\mathcal{P}_j$ denotes the set of buses in the \emph{unique} path from bus $j$ to bus 0 (including bus $j$ itself). \emph{Without loss of generality, we uniquely label a branch by its downstream end bus} (i.e., branch $j$'s downstream end is bus $j$). In this way, we are able to characterize the network only by bus labels.

Our proposed SM data-based topology and parameter identification methods build on the branch flow model \cite{ME1989} that relaxes the voltage angle. For notation convenience in later method development, we {modify} the original version by \emph{splitting} the power balance equations as,
\begin{subequations}\label{fullBFM}
\begin{align}
   & P_{j}=\sum_{k\in\mathcal{C}_j}\bar{P}_{k}-p_j,\,\,\bar{P}_j =P_j+r_{j}\cdot\left({P_j^2+Q_j^2}\right)/{v_j}\\
    & Q_{j} =\sum_{k\in\mathcal{C}_j}\bar{Q}_{k}-q_j,\,\,
    \bar{Q}_j =Q_j+x_{j}\cdot\left({P_j^2+Q_j^2}\right)/{v_j}\\
 & v_i-v_j =2\left(r_jP_j+x_jQ_j\right)+\left(r_j^2+x_j^2\right)\cdot\left({P_j^2+Q_j^2}\right)/{v_j}
\end{align} 
\end{subequations}
where $p_j,q_j$ denote the \emph{net} active/reactive power injection at bus $j$; $\bar{P}_j,\bar{Q}_j$ denote the real and reactive power flowing from the upstream bus $i$; ${P}_j,{Q}_j$ denote the real and reactive power flowing to the downstream bus $j$; $r_j,x_j>0$ are the line resistance and reactance; $v_i$ and $v_j$ are the \emph{squared} voltage magnitude at buses $i$ and $j$. 
\iffalse
\begin{figure}[t!]
  \centering
  \includegraphics[width=1.6
  in]{branchflow.png}\\
  \caption{Notational illustration of  branch flow model.}\label{branchflow}
  	\vspace{-1em}
\end{figure}
\fi

Different than the power flow analysis that solves voltages and line flows, the line impedances and grid connectivity will be extracted based on the known voltage and line flows, which is occasionally referred to as \emph{inverse power flow} in this paper \cite{yuan2020}.

To further reduce the solution space, we will leverage a library of conductor types existing in the network, as a constraint in our proposed optimization models. This information is usually available in practice, in the sense that the conductor types are often well recorded by utilities, although they may not know a specific line's conductor type. 
\section{Network Topology Identification via Weighted Laplacian Matrix}
In this section, we firstly develop an optimization-based topology identification method for distribution grids with the homogeneous R/X ratio, and then prove its robustness against the variability of R/X ratios.
\subsection{Link Between Grid Topology and Power Flow} 
\label{topology_1}
The proposed topology identification approach builds on the linear approximation of (\ref{fullBFM}) that neglects the nonlinear terms in (\ref{fullBFM}), which yields a (compact) voltage-power injection relationship as,
\begin{align}\label{vpq}
{\bf v}\simeq{2{\bf A}^{-T}{\bf R}{\bf A}^{-1}}{\bf p}+{2{\bf A}^{-T}{\bf X}{\bf A}^{-1}}{\bf q}{-{v}_0{\bf A}^{-T}{\bf a}_0}
\end{align}
where ${\bf v}:=[v_1,\ldots,v_n]^T$, ${\bf p}:=[p_1,\ldots,p_n]^T$, and ${\bf q}:=[q_1,\ldots,q_n]^T$ denote the vectors collecting squared bus voltage magnitudes, real power and reactive power injections at buses $1,...,n$, respectively; $[{\bf a}_0,{\bf A}^T]^T\in\{0,\pm1\}^{(n+1)\times n}$ is the incidence matrix of the radial-topology graph where ${\bf a}_0^T$ denotes the first row of the incidence matrix; ${\bf R}:={\rm diag}(r_1,...,r_n)$ and ${\bf X}:={\rm diag}(x_1,...,x_n)$ are diagonal matrices with $j$-th diagonal entry being the resistance and reactance of $j$-th branch.

\emph{Remark 1 (Sort Order within  $\bf p,q$ and $\bf v$):} It should be clarified that the entries within vectors $\bf p,q$ and $\bf v$ can be sorted without any prior restriction. More clearly, buses $1,...,n$ can be arbitrarily labelled regardless of the actual bus position in the network. The \emph {only} requirement is that they should be organized in a coherent way, meaning $p_j,q_j$ and $v_j$ that characterize bus $j$ should come from the same SM. 

For a radial distribution network, the reduce incidence matrix ${\bf A}:=[a_{ij}]_{n\times n}$ is non-singular \cite{BR2010} and ${\bf A}^{-T}{\bf a}_0=-{\bf 1}_n$. Therefore, a variant of (\ref{vpq}) reads,
\begin{align}\label{YB}
    \frac{1}{2}\underbrace{{\bf A}{\bf X}^{-1}{\bf A}^T}_{\bf Y}({\bf v}-v_0{\bf 1}_n)=\underbrace{{\bf A}{\bf X}^{-1}{\bf R}{\bf A}^{-1}}_{\bm\Phi}{\bf p}+{\bf q}
\end{align}
where $\bf Y$ is the weighted \emph{Laplacian} matrix of the network.

\emph{Proposition 1:} ${\bf Y}:=[y_{ij}]_{n\times n}$ is a sparse symmetric matrix with the entries being: 
\begin{align}\label{Ymatrix}
    y_{ij}=y_{ji}=\left\{\begin{matrix}
    -{1}/{x_j},&{\rm if}\,\,\, j\in\mathcal{C}_i \\
    \sum\limits_{k\in\{j\}\cup \mathcal{C}_j}{1}/{x_k},&{\rm if}\,\,\, i=j\\
    0,&\rm otherwise.
    \end{matrix}\right.
\end{align}

%\emph{Proof:} See Appendix. \qed
Mathematically, the rational behind $\bf Y$ is: for any two distinct buses $i$ and $j$, $y_{ij}<0$ if they are (directly) physically connected and otherwise, $y_{ij}=0$. \footnote{Furthermore, in a physical sense,  $\bf Y$ is structurally close to the admittance matrix  but without considering the line resistance.} Thus, if one can (approximately) identify $\bf Y$ that uniquely characterizes the connectivity, the topology can be extracted. This inspires our $\bf Y$-based topology identification.
\subsection{Identification Model}
We thus attempt to develop a regression model of $\bf Y$ based on (\ref{YB}) and the measurements of $\bf p,q,v$ and $v_0$ that can be obtained from SM data. It minimizes the mismatch between both sides of (\ref{YB}). Unfortunately, $\bm\Phi$ involves the network topology and parameters that are unknown yet. 

But interestingly, suppose the network has a homogeneous R/X ratio, i.e., \begin{align}
\frac{r_1}{x_1}=\cdots=\frac{r_n}{x_n}=\lambda,
\end{align} 
${\bm\Phi}$ reduces to
\begin{align}\label{Phi1}
    {\bm\Phi}={\bf A}{\rm diag}(r_1/x_1,...,r_n/x_n){\bf A}^{-1}={\lambda}{\bf I}_n
\end{align}
and accordingly, (\ref{YB}) boils down to,
\begin{align}
    {\bf Y}({\bf v}-v_0{\bf 1}_n)=2(\lambda{\bf p}+{\bf q})
\end{align}
which exactly eliminates the requirement of prior information regarding $\bf A$, $\bf R$ and $\bf X$, and purely relies on $\bf p, q, v$ and $v_0$.

Then, defining the mismatch vector regarding $k$-th sample,
\begin{align}\label{ek}
    {\bf e}^{(k)}:={\bf Y}\big({\bf v}^{(k)}-v_0^{(k)}{\bf 1}_n\big)-2\lambda{\bf p}^{(k)}-2{\bf q}^{(k)},\,\forall k
\end{align}
and ${\bf e}:=[({\bf e}^{(1)})^T,...,({\bf e}^{(K)})^T]^T$ where $K\gg n$ is the total number of samples,
a linear LS regression model of $\bf Y$ reads,
\begin{align}\label{T0}
\underset{{\bf Y},\lambda}{{\rm minimize}}\hspace{1.5mm}||{\bf e}||_2^2.
%\,\,\,{\rm subject}\,{\rm to}\hspace{1.5mm}{\bf Y}:\,\,\text{a symmetric matrix}.
\end{align}
\iffalse
with 
\begin{align}
 \nonumber f({\bf e}):=\left\{\hspace{-2mm}\begin{array}{l}
    ||{\bf e}||_1,\,\,\hspace{1.5mm} \text{for least-$\mathcal{L}_1$-norm model}\\
        ||{\bf e}||_\infty,\,\, \text{for least-$\mathcal{L}_\infty$-norm model}\\
    ||{\bf e}||_2,\,\,\hspace{1.5mm}\text{for least-$\mathcal{L}_2$-norm model}\\
        ||{\bf e}||^2_2,\,\,\hspace{1.5mm}\text{for least-square model.}\\
    \end{array}\right.
\end{align}
For the least-square model (termed as T3), the model is inherently a convex unconstrained QP program. The least $\mathcal{L}_1$-norm (T1) and $\mathcal{L}_\infty$-norm (T2) models can be reformulated as LP programs while the $\mathcal{L}_2$-norm (T4) based model can be rewritten as an SOCP program by introducing auxiliary variables. We refer the readers to Section IV for more details of the similar mathematical manipulations regarding $\mathcal{L}_1$-norm, $\mathcal{L}_\infty$-norm  and $\mathcal{L}_2$-norm operators. After that, all of the models at hand can be efficiently solved by off-the-shelf solvers.\fi

Clearly, this fitting model is an unconstrained QP program, of which a unique closed-form solution $(\bf Y^\star,\lambda^\star)$ can be obtained if it is strictly convex.

\subsection{Recovering Topology From Weighted Laplacian Matrix} \label{topology_2}
Recovering the topology from $\bf Y^\star$ can be cast as an \emph{anomaly detection} problem based on the property in Prop. 1. Considering the sparsity of the distribution system topology, a density-based spatial clustering of applications with noise method \cite{dbscan} is utilized here, which is tabulated as Algorithm \ref{dbscan_algorithm}. The rationale behind our task is that most of the entries in ${\bf Y}^\star_i$ for all $i$, are concentrated on a small range, which can be grouped into several clusters that represent the unconnected buses; the non-diagonal entries that do not belong to these clusters are declared as anomalies that indicate the connectivity. To achieve this, our method uses a minimum density level estimation based on two user-defined hyperparameters, a threshold for the minimum number of neighbors, $\gamma$, and the radius $\xi$. $y_{ij}^\star$ with more than $\gamma$ neighbors within $\xi$ distance are considered to be a core point. All neighbors within the $\xi$ radius of a core point are considered to be part of the same cluster as the core point. Based on multiple core points, all entries in $\bf Y^\star$ can be separated by clusters of lower density. The cluster with the minimum entries is considered to contain the connected buses. Overall, our method leverages the density drop between the unconnected and the connected entries in $\bf Y^\star$ to detect the cluster boundaries for recovering topology from estimated weighted Laplacian matrix. Unlike other clustering algorithms that assume normally shaped clusters, this method is capable of finding clusters with arbitrary shapes and sizes. Moreover, it does not require \emph{a priori} specification on the number of clusters, therefore, ensuring the robustness and practicality \cite{JS2016}. 

\begin{algorithm}[t]
\renewcommand\baselinestretch{1}\selectfont%\small
\caption{Recovering Topology From ${\bf Y}^\star$ by Clustering}
\begin{algorithmic}[0]\label{dbscan_algorithm}
\STATE \hspace{-3mm}{\bf Initialization}: Initialize $i\leftarrow 1,j\leftarrow 1$, $\gamma$, $\xi$
\STATE \hspace{-3mm}{\bf repeat}
\STATE [{\bf S1}]: Select the $i$th row of ${\bf Y}^\star$.
\STATE \hspace{-0mm}{\bf repeat}
\STATE \hspace{3mm}[{\bf S2}]: Pick $y^\star_{ij}$ and retrieve all direct density-reachable points using $\xi$.\\
%\hspace{11mm}from $y^\star_{ij}$.
\STATE \hspace{3mm}[{\bf S3}]: Based on $\gamma$, if $y^\star_{ij}$ is a core point, a cluster is formed; \\ \hspace{11mm}otherwise, update $j\leftarrow j+1$.  
\STATE \hspace{-0mm}{\bf until} {$j=n$} or no new point can be added to any cluster
\STATE [{\bf S4}]: Update $i\leftarrow i+1$.
\STATE \hspace{-3mm}{\bf until} {$i=n$}.
\end{algorithmic}
\end{algorithm}

%According to Prop. 1, a simple criterion is identifying the negative non-diagonal entries of $\bf Y$ with a large absolute value. Suppose $(\bf Y^\star, \lambda^\star)$ is the optimal solution of (\ref{T0}), an easy-to-implement method is setting a threshold $\varepsilon<0$, i.e.,
%\begin{align}
%Y_{ij}^{\star}\leq\varepsilon.
%\end{align}
%Many advanced data analytics methods for xxxx, e.g., xx, can be also used to filter the targeted entries.

\emph{Remark 2 (Robustness Against Inexact Estimation of $\bf Y$):} 
Given that the nonlinearity of power flow is dropped in the regression model, it is unlikely to solve $\bf Y^\star$ to be exactly equal to the true value of $\bf Y$ (denoted as $\bf Y^\ast$). Yet, it actually does \emph{not} require an exact estimation of $\bf Y$, because the topology is only sensitive to the structural feature rather than its exact value. This implies the connectivity behind two physical buses across a short line segment is easy to be identified. 
\begin{figure*}[t]
	\centering
	\begin{tikzpicture}
\begin{axis}[
xtick={1,2,...,10},
ylabel={$r_j\,\, (\rm 10^{-2} p.u.)$},
xmin=0.2,xmax=10.8,
legend style={at={(0.75,0.95)},
anchor=north,legend columns=-1},
ybar,bar width=4pt,
width=0.49*\textwidth,height=0.618*0.5*\textwidth,
legend columns=1,
xlabel={Branch No.}
]
\addplot[gray,fill=lightgray!30!white] table[y index = 1]{linearmodel_r.txt};
\addlegendentry{\small Benchmark};
\addplot[red,fill=red!30!white] table[y index = 3]{linearmodel_r.txt};
\addlegendentry{\small  With Library};
\addplot[blue,fill=blue!30!white] table[y index = 2]{linearmodel_r.txt};
\addlegendentry{\small Without Library};
\end{axis}
\end{tikzpicture}\hspace{3.2mm}
\begin{tikzpicture}
\begin{axis}[
xtick={1,2,...,10},ytick={0,0.5,1.0,1.5,2.0},
ylabel={$x_j\,\,(\rm 10^{-2} p.u.)$},
xmin=0.2,xmax=10.8,ymin=-0.4,ymax=2.05,
legend style={at={(0.75,0.95)},
anchor=north,legend columns=-1},
ybar,bar width=4pt,
width=0.49*\textwidth,height=0.618*0.5*\textwidth,
legend columns=1,
xlabel={Branch No.}
]
\addplot[gray,fill=lightgray!30!white] table[y index = 1]{linearmodel_x.txt};
\addlegendentry{\small Benchmark};
\addplot[red,fill=red!30!white] table[y index = 3]{linearmodel_x.txt};
\addlegendentry{\small With Library};
\addplot[blue,fill=blue!30!white] table[y index = 2]{linearmodel_x.txt};
\addlegendentry{\small Without Library};
\end{axis}
\end{tikzpicture}
	\caption{LS-based line parameter estimation results of the modified IEEE 13-bus test feeder (see Section V for details) based on the linearized inverse power flow model with and without the help of a R/X ratio library, where the  ``Benchmark" means the true values of $r_j$ and $x_j$.}
	\label{linearizedmodel}
\end{figure*}
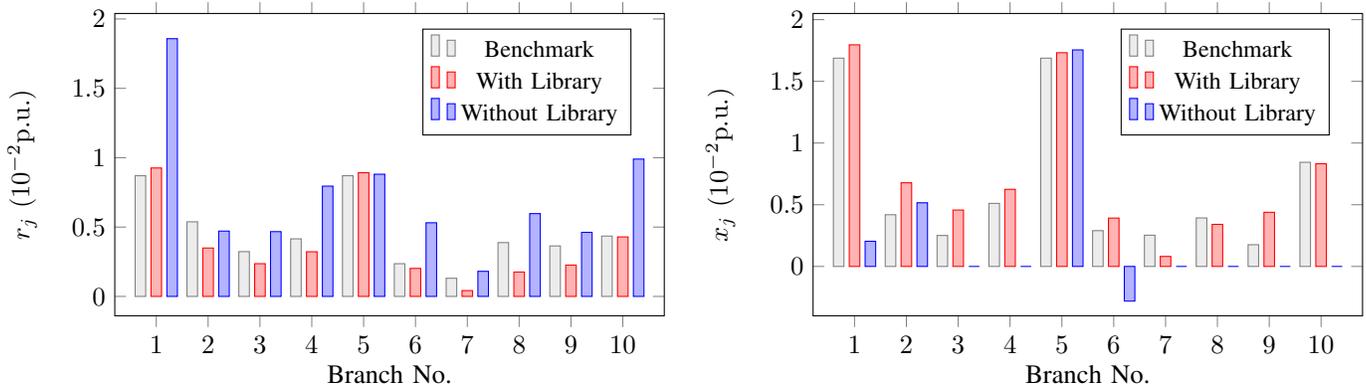
\subsection{Robustness Analysis on Heterogeneous Networks} \label{topology_3}
As mentioned above, the proposed regression model is derived on the assumption of a homogeneous R/X ratio [c.f. (6)], which may not be true in practical networks. However, it is widely believed to hold that a  distribution network under a given voltage level has the \emph{moderately heterogeneous} R/X ratios in practice. In what follows, we will prove that our proposed method is also robust on such cases.

Let $\lambda^\ast:=(\lambda_1+\cdots+\lambda_n)/n$ be the mean of R/X ratios and accordingly, let $
\lambda_j:=\lambda^\ast+\Delta\lambda_j,\,\forall j$
where $\Delta\lambda_j$ denotes the deviation to $\lambda$. Therefore, we have,
\begin{align}\label{Phi2}
    {\bm\Phi}={\lambda}^\ast{\bf I}_n+\bm\Delta_\lambda
\end{align}
where ${\bm\Delta}_\lambda:={\bf A}{\rm diag}(\Delta\lambda_1,\ldots,\Delta\lambda_n){\bf A}^{-1}$.

\emph{Proposition 2:} Matrix $\bm\Delta_\lambda:=[\Delta_{ij}]_{n\times n}$ is a matrix with the entries being,
\begin{align}\label{Deltamatrix}
    \Delta_{ij}=\left\{\begin{matrix}
    \Delta\lambda_i-\Delta\lambda_{\mathcal{C}_i\cap\mathcal{P}_j}, &{\rm if}\,\,\, i\in\mathcal{P}_j \\
    \Delta\lambda_i,&{\rm if}\,\,\,i=j\\
    0,&\rm otherwise.
    \end{matrix}\right.
\end{align}

Observe, $\bm\Delta_\lambda$ can quantify the heterogeneity of R/X across the whole network. For a relatively homogeneous network, $\lambda_j\simeq\lambda^\ast, \forall j$ and $|\Delta\lambda_i-\Delta\lambda_j|\simeq 0, \forall i,j$. Therefore, $\bm\Delta_\lambda$ will not significantly affect the solution of (\ref{T0}) provided it is numerically stable, and for a strictly homogeneous network, (\ref{Phi2}) completely reduces to (\ref{Phi1}).

Let ${\rm vec}:\mathbb{R}^{n\times n}\rightarrow\mathbb{R}^{n^2\times 1}$ be a mapping from a $n\times n$ symmetric matrix to a $n^2$-dimensional column vector, which vectorizes a matrix:
\begin{align}
   {\rm vec}({\bf Y}):=[y_{11},\ldots,y_{1n},\ldots,y_{n1},\ldots,y_{nn}]^T.
\end{align}
The following proposition further reveals the effect of heterogeneity of R/X ratios by quantifying the relationship between $\bm\Delta_\lambda$ and  $\bf Y^{\star}-\bf Y^\ast$. 

\emph{Proposition 3:}  Suppose 
(\ref{vpq}) holds and (\ref{T0}) is strictly convex, the distance between $\bf Y^\star$ and $\bf Y^\ast$ is bounded by,
\begin{align}
    \left\|{\rm vec}({\bf Y}^\star)-{\rm vec}({\bf Y^\ast})\right\|\leq\epsilon\left\|\bm\Delta_\lambda\right\|
\end{align}
where $\epsilon$ is a constant related to the sample data ${\bf v}^{(k)},{\bf p}^{(k)}$ and ${\bf q}^{(k)}$, $k=1,...,K$, which will be detailed in Appendix. 

%\emph{Proof:} See Appendix.
\section{Line Impedance Estimation: Convex Regression Models and Bottom-Up Sweep Framework}
We develop two types of regression models for impedance identification of a \emph{single} branch: (1) an LAD model with mixed-integer semidefinite programming (MISDP) formulation, and (2) an LS model with mixed-integer second-order cone programming (MISOCP) formulation, which are then embedded with a \emph{bottom-up} sweep algorithm to accomplish the parameter estimation across the entire network. 

Keep in mind that the proposed regression models will be built on full nonlinear inverse power flow instead of its linearized counterpart, in the sense that the latter may be unable to accurately recover the parameters especially when the regression problem is ill-posed; see Fig. \ref{linearizedmodel} for a numerical test on the IEEE 13-bus feeder. Note that, the non-convexity of nonlinear inverse power flow model makes the regression problems NP-hard even after continuous relaxation. This motivates us to specially address their convexifications as well.  
\subsection{Regression Models for A Single Branch}
The line impedance estimation establishes on the voltage drop relationship (\ref{fullBFM}e) over a branch. Define the vector of model mismatch ${\bf e}_{j}:=[e_j^{(1)},...,e_j^{(K)}]^T$ for all $j\in\mathcal{N}$ with 
\begin{align}\label{vareps}
\nonumber e^{(k)}_{j}:=&\hspace{1mm}v_i^{(k)}-v_j^{(k)}-2\big(r_jP_j^{(k)}+x_jQ_j^{(k)}\big)\\&-\left(R_j+X_j\right)\cdot\left[{\big(P_j^{(k)}\big)^2+\big(Q_j^{(k)}\big)^2}\right]/{v_j^{(k)}},
\,\forall k
\end{align}
where 
$R_j:=r_j^2$ and $X_j:=x_j^2$. 

The impedance estimation minimizes the $\mathcal{L}_1$-norm (LAD) or square (LS) of ${{\bf e}}_j$, which is expected to hold the following features. On one hand, the nonlinearity of the inverse power flow is well-captured to guarantee the estimation accuracy. On the other hand, the library of R/X ratios (obtained from the line conductor library) is exploited to significantly narrow the solution space; otherwise, the solution may easily fall into a remote local optima. Therefore, the line impedance estimation, which is inherently a combinatorial optimization problem, can be cast as a mixed-integer nonlinear programming (MINLP) model by introducing the binary variables $\alpha_1,..,\alpha_Z$,
\begin{subequations}\label{P0}
\begin{align}
\underset{\alpha_z,r_j,x_j,R_j,X_j}{{\rm minimize}}\hspace{2mm}&f({\bf e}_j)\\
{\rm subject}\,{\rm to}\hspace{2mm}
&R_j=r_j^2\\
&X_j=x_j^2\\
&r_j=\sum_{z=1}^{Z}\lambda_z\alpha_zx_j\\
&\sum_{z=1}^{Z}\alpha_z=1,\,\alpha_z\in\{0,1\},\, \forall z.
\end{align}
\end{subequations}
where $f({\bf e}_j):=||{\bf e}_j||_1$ for the LAD regression and $f({\bf e}_j):=||{\bf e}_j||^2_2$ for the LS regression, respectively.
\iffalse
\begin{align}
 \nonumber  f({\bf e}_j):=\left\{\hspace{-2mm}\begin{array}{ll}
    ||{\bf e}_j||_1,&\text{for least-$\mathcal{L}_1$-norm model}\\
 %       ||{\bf e}_j||_\infty,&\text{for least-$\mathcal{L}_\infty$-norm model}\\
%    ||{\bf e}_j||_2,\,\,\hspace{1.5mm}\text{for least-$\mathcal{L}_2$-norm model}\\
        ||{\bf e}_j||^2_2,&\text{for least-square model.}\\
    \end{array}\right.
\end{align}
\fi

The Big-M technique is exploited to linearize the bilinear term $\alpha_zx_j$ as,
\begin{align}\label{bigM}
    -M_j(1-\alpha_z)\leq r_j-\lambda_zx_j\leq M_j(1-\alpha_z),\,\, \forall z
\end{align}
where $M_j$ is a large real number.
\iffalse
\emph{Remark x (Selection of $M_j$):} The design of $M_j$ is usually tricky. Theoretically, it can be designed arbitrarily large. However, bearing in mind that if $M_j$ is too large, the resultant model may suffer from numerical instability. Hence, a theoretical lower bound of $M_j$ is considered here:
\begin{align}
    M_j\geq{\rm max}\big\{|r_j-\lambda_zx_j|,z=1,...,Z\big\}.
\end{align}
Obviously, this requires the prior information of $r_j,x_j$ which are ready to be solved. However, it is widely believed that utilities usually have some rough estimation of the length of distribution lines (such as some geographical information), so that $M_j$ can be determined in a rather rational way.
\fi

While (\ref{P0}) can be handled by some general MINLP solvers,
%\footnote{E.g., $\mathtt{KNITRO}$ or $\mathtt{BNB}$ where $\mathtt{BNB}$ should combine with a lower-level general NLP solver e.g., $\mathtt{IPOPT}$ or $\mathtt{fmincon}$.},
there is no guarantee of global optimality since its continuous relaxation counterpart is still non-convex due to the quadratic equalities (\ref{P0}b) and (\ref{P0}c). Therefore, in what follows, we will discuss the convexification techniques for the LAD and LS regression models, respectively.
\subsubsection{Least Absolute Deviations Regression Model}
To make the optimization model tractable, we  rewrite the cost function without $\mathcal{L}_1$-norm operator by introducing the auxiliary variables $\theta_j^{(1)},...,\theta_j^{(K)}$:
\begin{align}
   f(\theta_j^{(1)},\ldots,\theta_j^{(K)})= \sum_{k=1}^K\theta^{(k)}_{j}
\end{align}
with the additional constraints,
\begin{align}
    \theta^{(k)}_{j}\geq\hspace{0mm}e^{(k)}_{j},\,
-\theta^{(k)}_{j}\leq\hspace{0mm}e^{(k)}_{j},\,\forall k.
\end{align}
\iffalse
\begin{subequations}\label{P1}
\begin{align}
{\bf (P1)}:\underset{\alpha_z,r_j,x_j}{{\rm minimize}}\hspace{2mm}\sum_{k=1}^K&\hspace{1mm}\theta^{(k)}_{j}\\
{\rm subject}\,{\rm to}\hspace{2mm}\theta^{(k)}_{j}&\geq\hspace{0mm}e^{(k)}_{j}(r_j,x_j,R_j,X_j),\forall k\\
-\theta^{(k)}_{j}&\leq\hspace{0mm}e^{(k)}_{j}(r_j,x_j,R_j,X_j),\forall k\\
R_j&=\hspace{0mm}r_j^2\\
X_j&=\hspace{0mm}x_j^2\\
    r_j-\lambda_zx_j&\geq -M_j(1-\alpha_z),\, \forall z\\
    r_j-\lambda_zx_j&\leq M_j(1-\alpha_z),\, \forall z\\
\sum_{z=1}^{Z}\alpha_z&=\hspace{0mm}1,\,\,\alpha_z\in\{0,1\},\, \forall z.
\end{align}
\end{subequations}
\fi

To tackle the non-convex quadratic equalities (\ref{P0}b) and (\ref{P0}c), we propose to convexify them via leveraging the SDP relaxation. We first rewrite  (\ref{P0}b) and (\ref{P0}c) as,
 \begin{subequations}\label{SDPrelaxation}
 \begin{align}
 {\bf W}_{j}^r&:=
\begin{bmatrix}
1 &r_{j}\\
r_{j}& R_{j}
\end{bmatrix}\succeq0,\,
{\rm rank}\,\{{\bf W}_{j}^r\}=1,\,\forall j\\
 {\bf W}_{j}^x&:=
\begin{bmatrix}
1 &x_{j}\\
x_{j}& X_{j}
\end{bmatrix}\succeq0,\,
{\rm rank}\,\{{\bf W}_{j}^x\}=1,\,\forall j.
 \end{align}
\end{subequations}

By removing the rank-1 constraints in (\ref{SDPrelaxation}), a  MISDP model whose continuous relaxation is a convex SDP, is given by,
\begin{subequations}\label{P1}
\begin{align}
\underset{\alpha_z,r_j,x_j,R_j,X_j,{\theta}_j^{(k)}}{{\rm minimize}}\hspace{2mm}\sum_{k=1}^K&\hspace{1mm}\theta^{(k)}_{j}\\
{\rm subject}\,{\rm to}\hspace{7mm}\theta^{(k)}_{j}&\geq\hspace{0mm}e^{(k)}_{j},\forall k\\
-\theta^{(k)}_{j}&\leq\hspace{0mm}e^{(k)}_{j},\forall k\\
{\bf W}_{j}^r&\succeq0\\
{\bf W}_{j}^x&\succeq0\\
\text{(\ref{P0}e) and (\ref{bigM})}.\hspace{-12mm}&
\end{align}
\end{subequations}
\iffalse
\subsubsection{Least-$\mathcal{L}_\infty$-Norm Model} The least $\mathcal{L}_\infty$-norm model [labelled as (P2)] is similar to (\ref{PL1}) but with the differences: i) the cost function is replaced by $f(\theta_j)=\theta_j$ and ii) (\ref{PL1}b)--(\ref{PL1}c) are modified as $\theta_{j}\geq e^{(k)}_{j}, \theta_{j}\geq -e^{(k)}_{j}, \forall k$. 
\fi
\subsubsection{Least-Squares Model}
The cost function can be 
rewritten by introducing the auxiliary variable $\mu_j$:
\begin{align}
    f(\mu_j)=\mu_j\,\,\, \end{align}
and additionally imposing the constraint:
    \begin{align}
    \mu_{j}\geq \|{\bf e}_j\|_2^2.
\end{align}

Relaxing the quadratic equalities (\ref{P0}b) and (\ref{P0}c) into $R_j\geq r_j^2$ and $X_j\geq x_j^2$, we obtain a MISOCP model as,
\begin{subequations}\label{P2}
\begin{align}
\underset{\alpha_z,r_j,x_j,R_j,X_j,\mu_j}{{\rm minimize}}\hspace{1mm}\mu_j\hspace{14mm}&\\
{\rm subject}\,{\rm to}\hspace{5mm}
\left\|\begin{bmatrix}
\frac{\mu_j-1}{2}\\[1mm]
{\bf e}_j
\end{bmatrix}\right\|_2&\leq\dfrac{\mu_j+1}{2}\\
\left\|\begin{bmatrix}
\frac{R_j-1}{2}\\[1mm]
{r}_j
\end{bmatrix}\right\|_2&\leq\dfrac{R_j+1}{2}\\
    \left\|\begin{bmatrix}
\frac{X_j-1}{2}\\[1mm]
{x}_j
\end{bmatrix}\right\|_2&\leq\dfrac{X_j+1}{2}\\
\text{(\ref{P0}e) and (\ref{bigM})}&.
\end{align}
\end{subequations}
\iffalse
\subsubsection{Least-$\mathcal{L}_2$-Norm Model}
    For the least $\mathcal{L}_2$-norm model [termed as (P4)], the difference from (P3) lies in that  (\ref{PLS}b) is replaced with $\mu_j\leq||{\bf e}_j||_2$.
\fi

So far, once we have the knowledge of $P_j, Q_j$ and $v_j$,  (\ref{P1}) and (\ref{P2}) can be built, of which the continuous relaxations are convex, and can be handled by MISDP or MISOCP solvers. Besides, the proposed methods can be readily extended to account for $\mathcal{L}_\infty$-norm or $\mathcal{L}_2$-norm-based regression models.

\subsection{Bottom-Up Sweep Algorithm}
Clearly, the development of (\ref{P1}) or (\ref{P2}) requires the knowledge of voltage magnitude and line flow values. Unfortunately, due to the low coverage of line flow sensors, there are few line flow measurements available. Exceptions are the \emph{tail} branches since they have no further downstream neighbors and thus, the line flows physically equal the power injections at the leaf buses, which can be measured by SMs. Moreover, as per (\ref{fullBFM}a)--(\ref{fullBFM}b), \emph{the line flow over a given branch can be calculated, provided all of its neighboring downstream line flows have been known.} These facts motivate the design of a bottom-up  (a.k.a. leaf-to-root) sweep algorithm that manipulates the line flow and line impedance estimation in an alternating way. 
\iffalse
\begin{figure}[t]
	\centering
	\includegraphics[width=2.2in]{IEEE13_L2R.pdf}
	\caption{Schematic diagram of the bottom-up sweep algorithm in a tree-like distribution feeder.}
	\label{Bottemupsweep}
\end{figure}
\fi

We first partition a radial distribution network into multiple layers which are labeled as $1,...,D$ where $D$ is the maximum \emph{depth} [see Fig. \ref{ieee133769}(a) for an example with $D=4$]. Physically, ``bus $j$ belongs to layer $d$" means there are $d$ intermediate line segments in the path from bus $j$ to the root bus 0. The bottom-up sweep algorithm with the breadth-first search is then tabulated as Algorithm 2.

\begin{algorithm}[t]
\renewcommand\baselinestretch{1}\selectfont%\small
\caption{Bottom-Up Sweep Algorithm}
\begin{algorithmic}[0]\label{bottomup}
\STATE \hspace{-3mm}{\bf Initialization}: Initialize $d\leftarrow D$.
\STATE \hspace{-3mm}{\bf repeat}
\STATE [{\bf S1}]: Update the ${P}_j^{(k)}$ and ${Q}_j^{(k)}$ by (\ref{fullBFM}a) and (\ref{fullBFM}b) for all $k=1,...,K$ and $j$ in layer $d$.
\STATE [{\bf S2}]: Calculate $r_j, x_j$ of each line segment  by solving (\ref{P1}) or (\ref{P2}) for all $j$ in layer $d$.
\STATE [{\bf S3}]: Calculate $\bar{P}_j^{(k)}$ and $\bar{Q}_j^{(k)}$ as per (\ref{fullBFM}c) and (\ref{fullBFM}d) for all $k=1,...,K$ and $j$ in layer $d$.
\STATE [{\bf S4}]: Update $d\leftarrow d-1$.
\STATE \hspace{-3mm}{\bf until} {$d=0$}.
\end{algorithmic}
\end{algorithm}

\emph{Remark 3 (Scalability):} The proposed line parameter identification method has good scalability because the optimization is performed in a branch-wise manner. The computation burdens of  (\ref{P1}) and (\ref{P2}) are only related to the number of samples and the size of library. The sweep algorithm only requires very simple algebraic operations for line flow computation and thus scales well with the network size. 

\emph{Remark 4 (Error Propagation):}  Interestingly, the proposed algorithm is inherently robust given that the estimation errors regarding downstream branches will only affect the line losses, which slightly contributes to the upstream-end line flows. It is therefore expected that the effects of errors can asymptotically diminish. This will be verified in the numerical tests. 
\section{Numerical Results}
\begin{figure*}[t]
	\centering
	\includegraphics[width=6.8in]{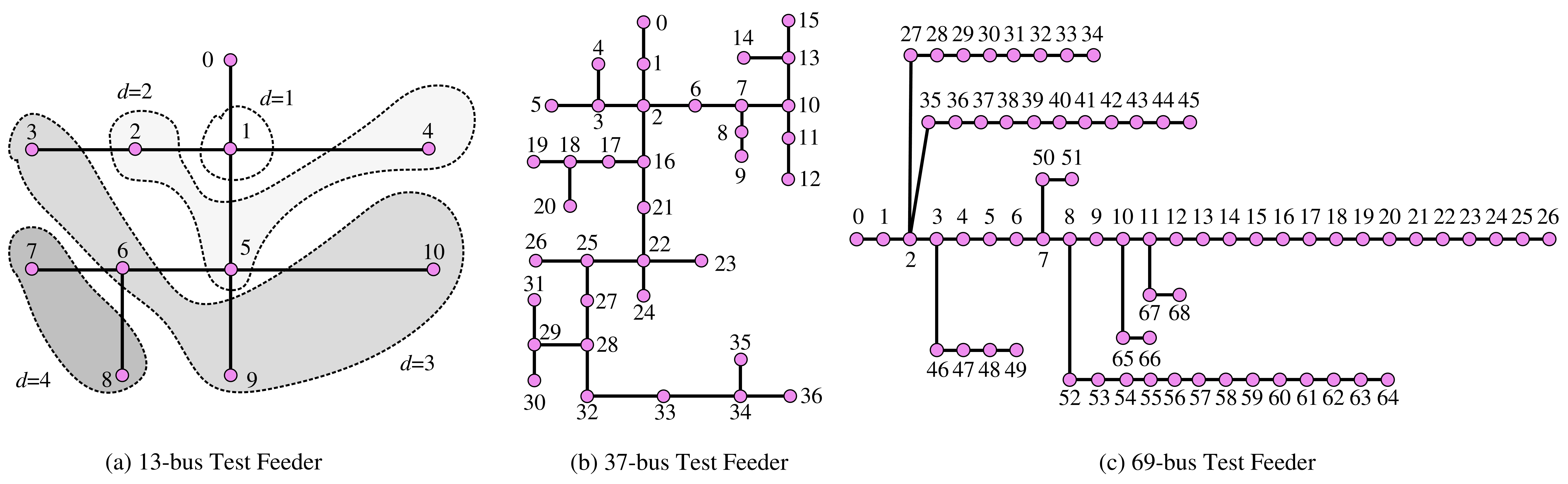}
	\caption{One-line diagrams of the modified IEEE (a) 13-bus, (b) 37-bus and (c) 69-bus test feeders (balanced) where the original 13-bus test feeder is modified to a 11-bus test feeder by removing the dummy buses 634 and 692 of the original case and the line impedance of 69-bus feeder are slightly modified to achieve several typical R/X ratios. The resultant R/X ratio libraries are $\{0.5153,1.2840,0.8124,0.8112,0.9864,2.0655\}$, $\{1.4536,1.6222,2.7482,1.9691\}$, and $\{0.4000,0.8000,0.9000,2.0000,2.9000,3.0000,3.1000,3.3000,3.4000\}$ in the three cases, respectively.}
	\label{ieee133769}
		\vspace{-1em}
\end{figure*}
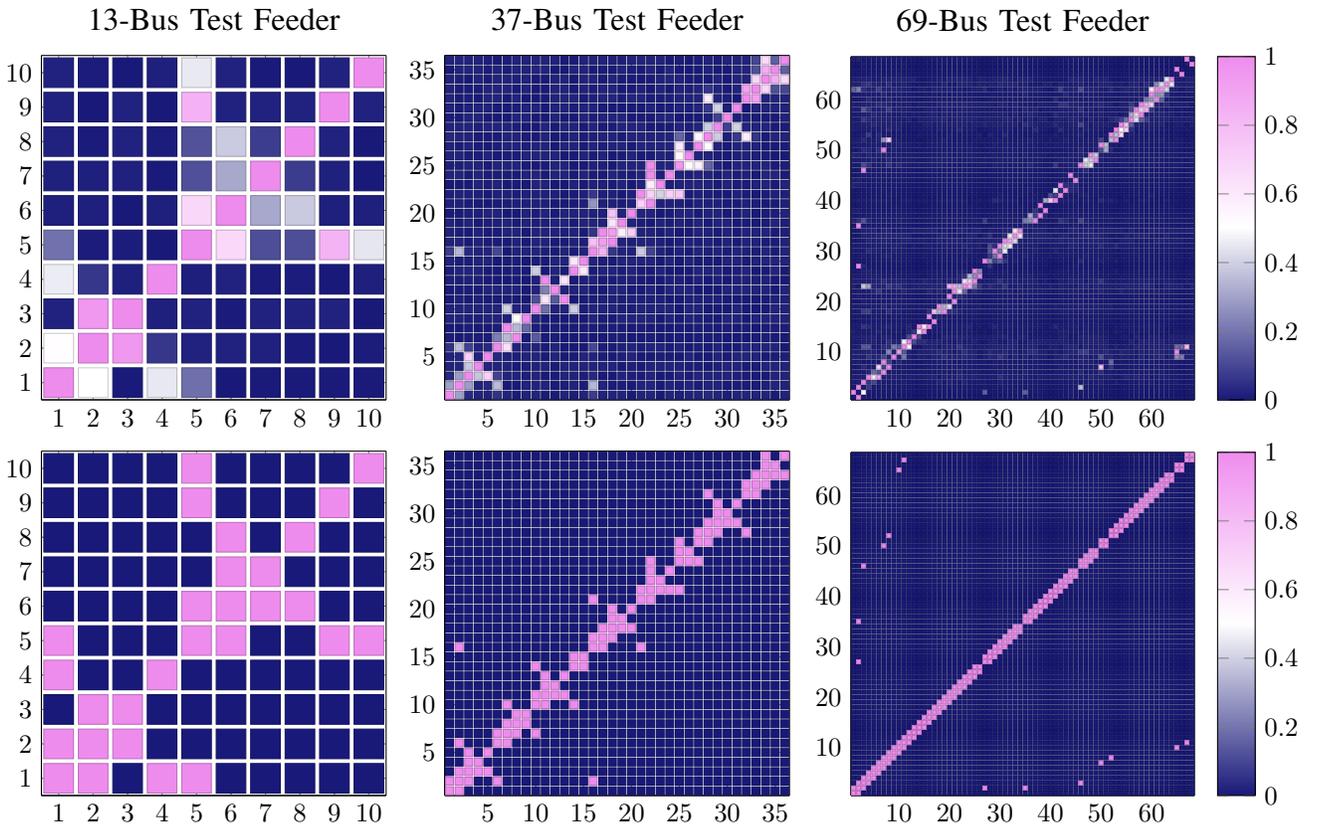
\begin{figure*}[t]
	\centering
\begin{tikzpicture}
  \begin{axis}[xtick={1,2,...,10},ytick={1,...,10},xmin=0.5,xmax=10.5,ymin=0.5,ymax=10.5,colormap/violet,view={0}{90},width=0.34*\textwidth,height=0.34*\textwidth,title={\large 13-Bus Test Feeder}]%
   \addplot3[only marks,scatter,mark=square*,mark size=5.6]file{Ybus13.txt};
\end{axis}
\end{tikzpicture}
\begin{tikzpicture}
  \begin{axis}[xtick={5,10,...,35},ytick={5,10,...,35},xmin=0.5,xmax=36.5,ymin=0.5,ymax=36.5,colormap/violet,view={0}{90},width=0.34*\textwidth,height=0.34*\textwidth,title={\large 37-Bus Test Feeder}]%
   \addplot3[only marks,scatter,mark=square*,mark size=1.5]file{Ybus37.txt};
\end{axis}
\end{tikzpicture}
\begin{tikzpicture}
  \begin{axis}[xtick={10,20,...,60},ytick={10,20,...,60},xmin=0.5,xmax=68.5,ymin=0.5,ymax=68.5,colormap/violet,view={0}{90},colorbar,width=0.34*\textwidth,height=0.34*\textwidth,title={\large 69-Bus Test Feeder}]%
   \addplot3[only marks,scatter,mark=square*,mark size=0.75]file{Ybus69.txt};
\end{axis}
\end{tikzpicture}

\begin{tikzpicture}
\begin{axis}[xtick={1,2,...,10},ytick={1,...,10},xmin=0.5,xmax=10.5,ymin=0.5,ymax=10.5,colormap/violet,view={0}{90},width=0.34*\textwidth,height=0.34*\textwidth
]%
\addplot3[only marks,scatter,mark=square*,mark size=5.6]file{13bus_Alg1.txt};
\end{axis}
\end{tikzpicture}
\begin{tikzpicture}
  \begin{axis}[xtick={5,10,...,35},ytick={5,10,...,35},xmin=0.5,xmax=36.5,ymin=0.5,ymax=36.5,colormap/violet,view={0}{90},width=0.34*\textwidth,height=0.34*\textwidth]%
   \addplot3[only marks,scatter,mark=square*,mark size=1.5]file{37bus_Alg1.txt};
\end{axis}
\end{tikzpicture}
\begin{tikzpicture}
  \begin{axis}[xtick={10,20,...,60},ytick={10,20,...,60},xmin=0.5,xmax=68.5,ymin=0.5,ymax=68.5,colormap/violet,view={0}{90},colorbar,width=0.34*\textwidth,height=0.34*\textwidth]%
   \addplot3[only marks,scatter,mark=square*,mark size=0.75]file{69bus_Alg1.txt};
\end{axis}
\end{tikzpicture}
	\caption{Results of topology identification of the modified IEEE 13-bus (left), 37-bus (middle), and 69-bus (right) test feeders. The top part shows the normalized counterpart of ${\bf Y}^\star$ obtained by model (\ref{T0}). The bottom part shows the output of  Algorithm 1, which represents the connectivity where ``0" denotes         ``unconnected" and ``1" denotes ``connected''.}
	\label{fig:topology_result}
		\vspace{-1em}
\end{figure*}
\begin{figure*}[t!]
	\centering
		\begin{tikzpicture}
\begin{axis}[title={\large 13-Bus Test Feeder},
xtick={1,2,...,10},
ylabel={$r_j\,\, (\rm 10^{-2} p.u.)$},
xmin=0.2,xmax=10.8,ymax=1.5,ymin=-0.03,
legend style={at={(0.68,0.95)},
anchor=north,legend columns=3},
ybar,bar width=4pt,
width=0.47*\textwidth,height=0.5*0.47*\textwidth,
legend columns=3,
xlabel={Branch No.}
]
\addplot[gray,fill=lightgray!30!white] table[y index = 1]{R13.txt};
\addlegendentry{\small Benchmark};
\addplot[red,fill=red!30!white] table[y index = 3]{R13.txt};
\addlegendentry{\small LAD};
\addplot[blue,fill=blue!30!white] table[y index = 2]{R13.txt};
\addlegendentry{\small LS};
\end{axis}
\end{tikzpicture}\hspace{2mm}
\begin{tikzpicture}
\begin{axis}[title={\large 13-Bus Test Feeder},
xtick={1,2,...,10},ytick={0,0.5,1.0,1.5,2.0},
ylabel={$x_j\,\,(\rm 10^{-2} p.u.)$},
xmin=0.2,xmax=10.8,ymin=-0.06,ymax=2.5,
legend style={at={(0.68,0.95)},
anchor=north,legend columns=-1},
ybar,bar width=4pt,
width=0.47*\textwidth,height=0.5*0.47*\textwidth,
xlabel={Branch No.}
]
\addplot[gray,fill=lightgray!30!white] table[y index = 1]{X13.txt};
\addlegendentry{\small Benchmark};
\addplot[red,fill=red!30!white] table[y index = 3]{X13.txt};
\addlegendentry{\small LAD};
\addplot[blue,fill=blue!30!white] table[y index = 2]{X13.txt};
\addlegendentry{\small LS};
\end{axis}
\end{tikzpicture}

\begin{tikzpicture}
\begin{axis}[title={\large 37-Bus Test Feeder},
xtick={1,2,...,36},xlabel={},ytick={0,0.5,1,1.5,2},
ylabel={$r_j\,\, (\rm 10^{-2} p.u.)$},
xmin=0,xmax=37,ymax=2.5,ymin=-0.1,
legend style={at={(0.145,0.96)},
anchor=north,legend columns=3},
ybar=.05cm,bar width=2pt,
width=0.95*\textwidth,height=0.25*0.95*\textwidth,
legend columns=3,
]
\addplot[gray,fill=lightgray!30!white] table[y index = 1]{R37.txt};
\addlegendentry{\small Benchmark};
\addplot[red,fill=red!30!white] table[y index = 3]{R37.txt};
\addlegendentry{\small LAD};
\addplot[blue,fill=blue!30!white] table[y index = 2]{R37.txt};
\addlegendentry{\small LS};
\end{axis}
\end{tikzpicture}

\begin{tikzpicture}
\begin{axis}[
xtick={1,2,...,36},
ylabel={$x_j\,\,(\rm 10^{-2} p.u.)$},
xmin=0,xmax=37,ymin=-0.05,ymax=0.9,
legend style={at={(0.145,0.96)},
anchor=north},
ybar=.05cm,bar width=2pt,
width=0.95*\textwidth,height=0.25*0.95*\textwidth,
legend columns=3,
xlabel={Branch No.}
]
\addplot[gray,fill=lightgray!30!white] table[y index = 1]{X37.txt};
\addlegendentry{\small Benchmark};
\addplot[red,fill=red!30!white] table[y index = 3]{X37.txt};
\addlegendentry{\small LAD};
\addplot[blue,fill=blue!30!white] table[y index = 2]{X37.txt};
\addlegendentry{\small LS};
\end{axis}
\end{tikzpicture}	

\begin{tikzpicture}
\begin{axis}[title={\large 69-Bus Test Feeder},
ymode=log,
ylabel={$r_j\,\, (\rm 10^{-2} p.u.)$},
xmin=0,xmax=69,ymax=1.5,xtick={5,10,...,65},
legend style={at={(0.135,0.96)},
anchor=north,legend columns=3},
ybar=.03cm,bar width=1pt,
width=0.95*\textwidth,height=0.25*0.95*\textwidth,
max space between ticks=20
]
\addplot[gray,fill=lightgray!30!white] table[y index = 1]{R69.txt};
\addlegendentry{\small Benchmark};
\addplot[red,fill=red!30!white] table[y index = 3]{R69.txt};
\addlegendentry{\small LAD};
\addplot[blue,fill=blue!30!white] table[y index = 2]{R69.txt};
\addlegendentry{\small LS};
\end{axis}
\end{tikzpicture}

\begin{tikzpicture}
\begin{semilogyaxis}[
ylabel={$x_j\,\,(\rm 10^{-2} p.u.)$},
xmin=0,xmax=69,ymax=1.5,xtick={5,10,...,65},
legend style={at={(0.135,0.96)},
anchor=north,legend columns=3},
ybar=.03cm,bar width=1pt,
width=0.95*\textwidth,height=0.25*0.95*\textwidth,xlabel={Branch No.}
]
\addplot[gray,fill=lightgray!30!white] table[y index = 1]{X69.txt};
\addlegendentry{\small Benchmark};
\addplot[red,fill=red!30!white] table[y index = 3]{X69.txt};
\addlegendentry{\small LAD};
\addplot[blue,fill=blue!30!white] table[y index = 2]{X69.txt};
\addlegendentry{\small LS};
\end{semilogyaxis}
\end{tikzpicture}	
	\caption{Results of line parameter estimation of 13-, 37- and 69-bus test feeders.}
	\label{RX69com}
		\vspace{-1em}
\end{figure*}
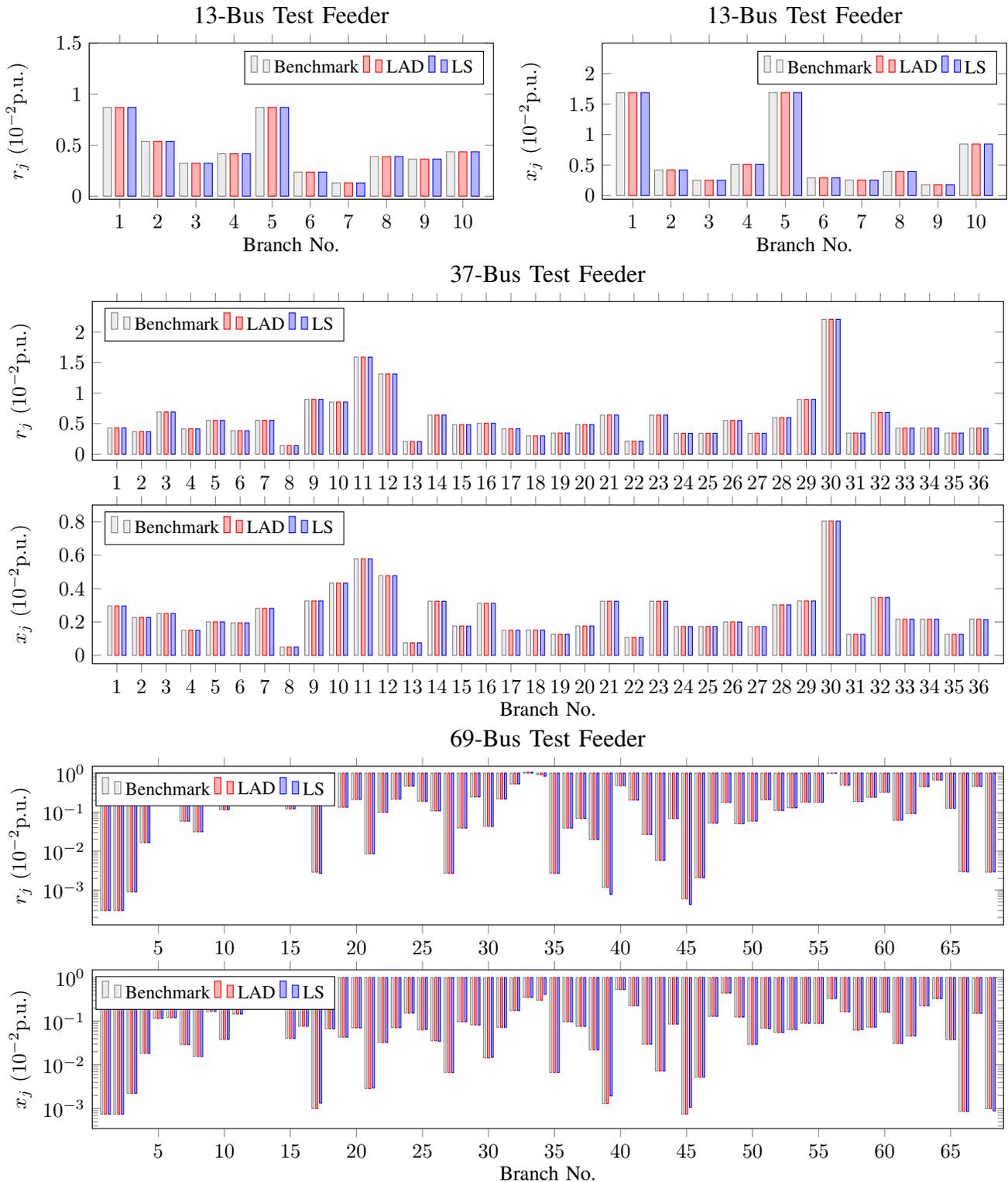
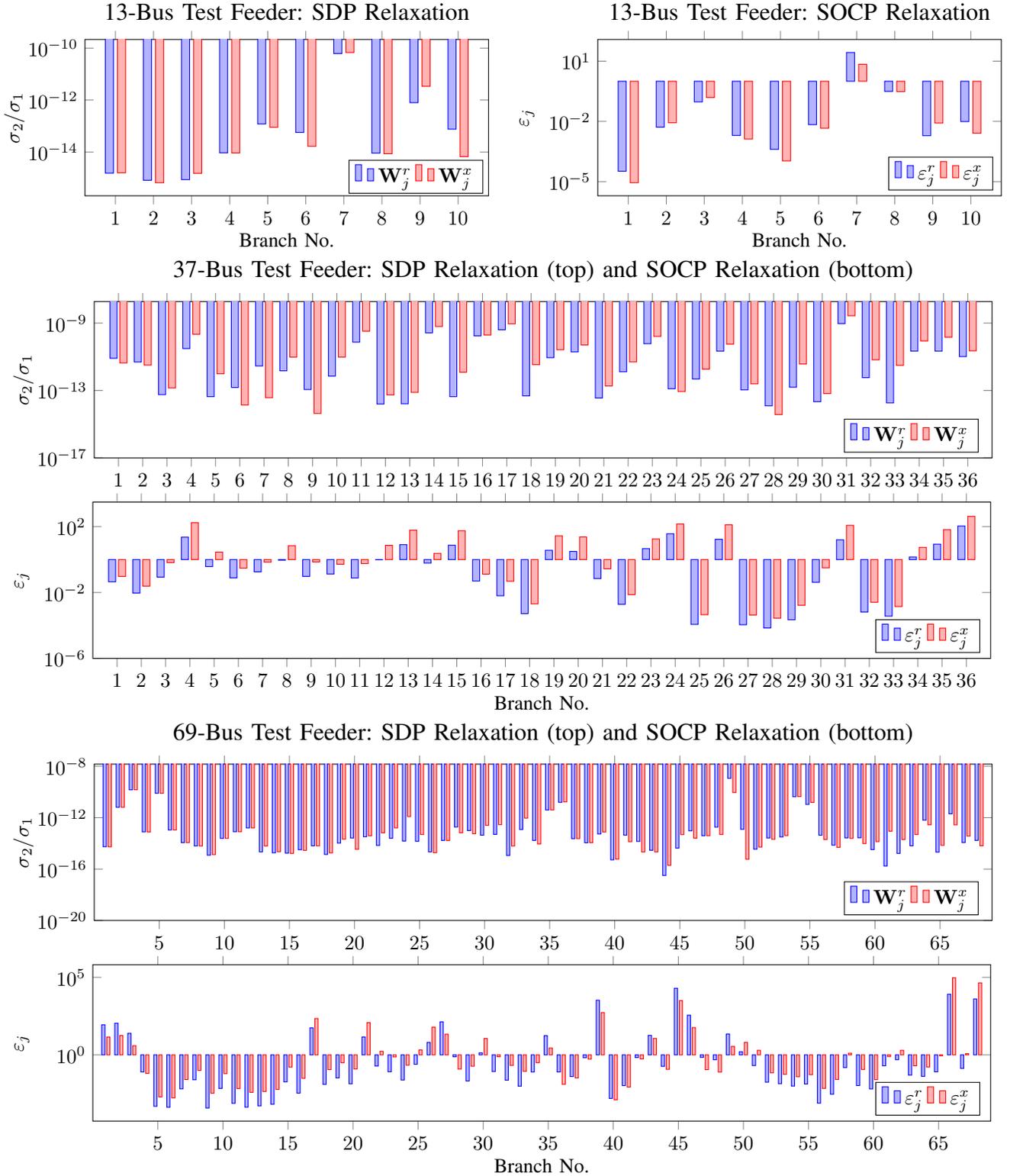
\begin{figure*}[t!]
	\centering
		\begin{tikzpicture}
\begin{semilogyaxis}[title={\large 13-Bus Test Feeder: SDP Relaxation},
xtick={1,2,...,10},
ylabel={$\sigma_2/\sigma_1$},
xmin=0.2,xmax=10.8,
legend style={at={(0.82,0.23)},
anchor=north,legend columns=3},
ybar,bar width=4pt,
width=0.475*\textwidth,height=0.5*0.475*\textwidth,
legend columns=2,
xlabel={Branch No.}
]
\addplot[blue,fill=blue!30!white] table[y index = 2]{error_sdp13.txt};
\addlegendentry{${\bf W}^r_j$};
\addplot[red,fill=red!30!white] table[y index = 1]{error_sdp13.txt};
\addlegendentry{${\bf W}^x_j$};
\end{semilogyaxis}
\end{tikzpicture}\hspace{3mm}
\begin{tikzpicture}
\begin{semilogyaxis}[title={\large 13-Bus Test Feeder: SOCP Relaxation},
xtick={1,2,...,10},
ylabel={$\varepsilon_j$},
xmin=0.2,xmax=10.8,
legend style={at={(0.85,0.25)},
anchor=north,legend columns=-1},
ybar,bar width=4pt,
width=0.475*\textwidth,height=0.5*0.475*\textwidth,
legend columns=2,
xlabel={Branch No.}
]
\addplot[blue,fill=blue!30!white] table[y index = 2]{error_socp13.txt};
\addlegendentry{$\varepsilon^r_j$};
\addplot[red,fill=red!30!white] table[y index = 1]{error_socp13.txt};
\addlegendentry{$\varepsilon^x_j$};
\end{semilogyaxis}
\end{tikzpicture}

\begin{tikzpicture}
\begin{semilogyaxis}[title={\large 37-Bus Test Feeder: SDP Relaxation (top) and SOCP Relaxation (bottom)},
xtick={1,2,...,36},xmin=0,xmax=37,ymin=1e-17,
ylabel={$\sigma_2/\sigma_1$},
legend style={at={(0.91,0.25)},
anchor=north,legend columns=2},
ybar=.05cm,bar width=3.5pt,
width=0.95*\textwidth,height=0.25*0.95*\textwidth,
legend columns=3,
]
\addplot[blue,fill=blue!30!white] table[y index = 2]{error_sdp37.txt};
\addlegendentry{${\bf W}^r_j$};
\addplot[red,fill=red!30!white] table[y index = 1]{error_sdp37.txt};
\addlegendentry{${\bf W}^x_j$};
\end{semilogyaxis}
\end{tikzpicture}

\begin{tikzpicture}
\begin{semilogyaxis}[
xtick={1,2,...,36},xmin=0,xmax=37,ymin=1e-6,
ylabel={$\varepsilon_j$},
legend style={at={(0.93,0.25)},
anchor=north},
ybar=.05cm,bar width=3.5pt,
width=0.95*\textwidth,height=0.25*0.95*\textwidth,
legend columns=2,
xlabel={Branch No.}
]
\addplot[blue,fill=blue!30!white] table[y index = 2]{error_socp37.txt};
\addlegendentry{$\varepsilon^r_j$};
\addplot[red,fill=red!30!white] table[y index = 1]{error_socp37.txt};
\addlegendentry{$\varepsilon^x_j$};
\end{semilogyaxis}
\end{tikzpicture}	

\begin{tikzpicture}
\begin{semilogyaxis}[title={\large 69-Bus Test Feeder: SDP Relaxation (top) and SOCP Relaxation (bottom)},
ylabel={$\sigma_2/\sigma_1$},
xmin=0,xmax=69,xtick={5,10,...,65},ymin=1e-20,
legend style={at={(0.91,0.25)},
anchor=north,legend columns=2},
ybar=.03cm,bar width=1.6pt,
width=0.95*\textwidth,height=0.25*0.95*\textwidth,
max space between ticks=20
]
\addplot[blue,fill=blue!30!white] table[y index = 2]{error_sdp69.txt};
\addlegendentry{${\bf W}^r_j$};
\addplot[red,fill=red!30!white] table[y index = 1]{error_sdp69.txt};
\addlegendentry{${\bf W}^x_j$};
\end{semilogyaxis}
\end{tikzpicture}

\begin{tikzpicture}
\begin{semilogyaxis}[
ylabel={$\varepsilon_j$},
xmin=0,xmax=69,xtick={5,10,...,65},
legend style={at={(0.93,0.25)},
anchor=north,legend columns=2},
ybar=.03cm,bar width=1.6pt,
width=0.95*\textwidth,height=0.25*0.95*\textwidth,xlabel={Branch No.}
]
\addplot[blue,fill=blue!30!white] table[y index = 2]{error_socp69.txt};
\addlegendentry{$\varepsilon^r_j$};
\addplot[red,fill=red!30!white] table[y index = 1]{error_socp69.txt};
\addlegendentry{$\varepsilon^x_j$};
\end{semilogyaxis}
\end{tikzpicture}	
	\caption{Exactness of SDP and SOCP relaxation of IEEE 13-, 37- and 69-bus test feeders.}
	\label{error133769}
		\vspace{-1em}
\end{figure*}
In this section, the proposed topology and parameter identification methods are verified on the modified IEEE 13, 37 and 69-bus test feeders, which are depicted in Fig. \ref{ieee133769}. The power flow analysis takes as input these distribution system models and the nodal load demand time-series with 1-h resolution and totally 200 samples, obtained by aggregating the load of a certain number of customers from a utility in Midwest, U.S. The computed voltages are treated as the voltage measurements. The optimization programs are solved by YALMIP Toolbox in MATLAB, along with the solver $\mathtt{MOSEK}$  \cite{YALMIP}.

\subsection{Results of Topology Identification}
The topology identification results of the modified IEEE 13-bus, 37-bus and 69-bus test feeders are depicted in Fig. \ref{fig:topology_result}. For data visualization, the min-max normalization  \cite{Goodfellow2016} is utilized to rescale the entries of ${\bf Y}^\star_i$ to be within [0,1] for all $i$. 

%\footnote{Min-max normalization is one of the most common ways to rescale data. For each row of the $|{\bf Y}^\star|$, the minimum value is converted into 0 and the maximum value is converted into 1. The remaining values are converted into a decimal between 0 and 1. }
The left-hand part of Fig. \ref{fig:topology_result} illustrates matrices $\bf Y^\star$ of each test feeder solved by the proposed model (\ref{T0}), which are the sparse symmetric matrices. Then, by using Algorithm 1, the estimated Laplacian matrix of each test feeder is obtained, as shown in the right-hand part of Fig. \ref{fig:topology_result}. The performance is validated by comparing the estimated connectivity and the real connectivity. In this work, the proposed method achieves an accuracy of topology recovery under all the three distribution feeders. Note that, this  verifies our method's robustness against heterogeneous R/X ratios. 
\subsection{Results of Line Parameter Estimation}
The line parameter estimation results of the modified IEEE 13-bus, 37-bus and 69-bus test feeders are depicted in Fig. \ref{RX69com}. As can be seen from Fig. \ref{RX69com},  the SDP-based LAD model precisely recovers the line impedance of each branch, under all the three test cases. 
In terms of $r_j$, the largest relative errors (among all branches) are $3.33\times10^{-5}\%$, $3.40\times10^{-4}\%$ and $1.44\times10^{-4}\%$ for the modified IEEE 13-bus, 37-bus and 69-bus test feeders, respectively; and as for $x_j$, the largest relative errors are $3.33\times10^{-5}\%$, $3.40\times10^{-4}\%$ and $7.06\times10^{-5}\%$, respectively.

The SOCP-based LS model works well with the modified IEEE 13-bus and 37-bus test feeders, with the largest relative errors $0.25\%$ and $0.95\%$ regarding $r_j$, and $0.25\%$ and $0.95\%$ regarding $x_j$, respectively. However, for the 69-bus test feeder, it fails to accurately identify the impedance on some branches, with the largest relative errors of $33.95\%$ and $46.81\%$ regarding $r_j$ and $x_j$, respectively, but such \emph{considerable} mis-estimation (i.e., relative error $\geq5\%$) only happens to a limited number of branches (17, 34, 39, 45 and 68). And we note that, the solutions on these branches fail to pick the right R/X ratio. However, it is observed that the mis-estimation does not much deteriorate the subsequent estimation of the upstream branches. This demonstrates the robustness against error propagation, as elaborated in Remark 4.

To quantify the exactness of SDP relaxation in (\ref{P1}), that is how close are the matrices ${\bf W}^{r}_j$ and ${\bf W}^x_j$ to rank one, one can compute the ration between their largest two eigenvalues, i.e., $\sigma_2\{\ast\}/\sigma_1\{\ast\}$. Similarly, as for the SOCP relaxation in (\ref{P2}), we quantify its exactness by the resultant errors $\varepsilon^r_j:=|(R_j-r_j^2)/r_j^2|$ and 
$\varepsilon^x_j:=|(X_j-x_j^2)/x_j^2|$. The results are depicted in Fig. \ref{error133769}. It is observed that the SDP relaxation is exact on all branches with $\sigma_2({\bf W}^{r}_j)/\sigma_1({\bf W}^{r}_j)$ and $\sigma_2({\bf W}^{x}_j)/\sigma_1({\bf W}^{x}_j)$ less than $10^{-8}$ in the three cases. In comparison, the SOCP relaxation is not well exact on some branches in the three cases; but as presented before, the estimation results show acceptable accuracy except several branches in the 69-bus case. 
\section{Conclusions}\label{conclusion}
In this paper, we have proposed a data-driven framework to accurately and efficiently extract distribution grid models from SM data. The proposed topology identification establishes on reconstructing the weighted Laplacian matrix of a homogeneous distribution circuit, which also exhibits provable robustness against heterogeneous R/X ratios. The nonlinear LAD and LS regression models for parameter identification are developed and convexified. They are capable of recovering the line impedances; but, in comparison, the LAD model with the SDP relaxation shows better exactness and stability.

This paper focuses on the balanced radial distribution grids with a full coverage of SMs. The unbalanced and meshed cases require further investigation; and moreover, the partial observability with imprecise measurement also need to be addressed in future work.

\section*{Appendix}
\subsection{Proof of Proposition 1} We have
\begin{align}
\nonumber  y_{ij}&={\bf A}_{i}{\bf X}^{-1}({{\bf A}^T})^{j}={\bf A}_{i}{\bf X}^{-1}{{\bf A}}_j^T\\&
    =\sum_{k=1}^n a_{ik}x_{k}^{-1}a_{jk}=\sum_{k=1}^n a_{jk}x_{k}^{-1}a_{ik}=y_{ji}.
\end{align}
Interestingly, if $i\neq j$, $a_{ik}\cdot a_{jk}=-1$ holds for $k=j$ and 
$a_{ik}\cdot a_{jk}=0$, otherwise; if $i=j$, $a_{ik}\cdot a_{jk}=1$ always holds for any branch $k$ with either end bus being $j$ and otherwise, $a_{ik}\cdot a_{jk}=0$. Accordingly, (\ref{Ymatrix}) follows. \qed
\subsection{Proof of Proposition 2} 
Let ${\bf B}:={\bf A}^{-1}$. First, as per the linear algebra theory, ${\bf I}_n\to{\bf B}$ can be accomplished via the elementary row operations, by which, in turn, one can exactly achieve ${\bf A}\to{\bf I}_n$. Second, given that $\bf A$ is the reduced incidence matrix of a tree-topology network, we have $a_{ij}=-1$ if $i=j$, $a_{ij}=1$ if $i\in\mathcal{P}_j\backslash\{j\}$ and otherwise, $a_{ij}=0$. To achieve ${\bf A}\to{\bf I}_n$, one has to add the rows $j$ for all $j\in\{j|i\in\mathcal{P}_j\}$ onto the row $i$, and then multiply row $i$ by $-1$. Accordingly, one can obtain ${\bf B}:=[b_{ij}]_{n\times n}$ with
\begin{align}\label{Bdef}
    {b}_{ij}=\left\{\begin{matrix}
    -1,&{\rm if}\,\,i\in\mathcal{P}_j\,\,{\rm or}\,\,i=j\\
    0,&\rm otherwise.
    \end{matrix}\right.
\end{align}
And then, we have
\begin{align}
    a_{ik}\cdot b_{kj}=\left\{
    \begin{matrix}
    1,&{\rm if}\,\,k\in\{i\}\cap\mathcal{P}_j\\
    -1,&{\rm if}\,\,k\in\mathcal{C}_i\cap\mathcal{P}_j\\
    0,&{\rm otherwise}
    \end{matrix}\right.
\end{align}
which indicates it is non-zero if and only if $i\in\mathcal{P}_j$.
Therefore,
\begin{align}
\nonumber
{\delta}_{ij}&={\bf A}_i{\rm diag}(\Delta\lambda_1,\ldots,\Delta\lambda_n){\bf B}^j=\sum_{k=1}^{n}a_{ik}\Delta\lambda_{k}b_{kj}\\
&=\left\{\begin{matrix}
    \Delta\lambda_i-\Delta\lambda_{\mathcal{C}_i\cap\mathcal{P}_j}, &{\rm if}\,\,\, i\in\mathcal{P}_j\backslash\{j\} \\
    \Delta\lambda_i,&{\rm if}\,\,\,i=j\\
    0,&\rm otherwise
    \end{matrix}\right.
\end{align}
for any $i,j\in\mathcal{N}$.\qed
\subsection{Proof of Proposition 3}
\iffalse
First, rearrange the lower triangular parts of $\bf Y$ to be a $n(n+1)/2$-dimensional column vector ${\bf u}$ with the entries being,
\begin{align}
%  {\bf y}^{\star}&:=[y_{11}^\star,...,y_{1n}^\star,y_{22}^\star,...,y_{2n}^\star,...,y_{nn}^\star]^T\\
% {\bf y}^\ast&:=[y_{11}^\ast,...,y_{1n}^\ast,y_{22}^\ast,...,y_{2n}^\ast,...,y_{nn}^\ast]^T.
u_m:=y_{ij},\,\, {\rm with}\,\,\,m=\frac{(i-1)\times i}{2}+j
\end{align}
for $i=1,\ldots,n$ and $j=1,\ldots,i$.
\fi
For notational simplicity, let $\bf y:= {\rm vec}(Y)$ and correspondingly, $\bf y^\star:= {\rm vec}(Y^\star)$ and $\bf y^\ast:={\rm vec}(Y^\ast)$.

Accordingly, we can replace the $\bf Y$-related term in (\ref{ek}) by,
\begin{align}
{\bf G}^{(k)}{\bf y}={\bf Y}\big({\bf v}^{(k)}-v_0^{(k)}{\bf 1}_n\big)
\end{align}
with 
\begin{align}
    {\bf G}^{(k)}:={\bf I}_n\otimes\big({\bf v}^{(k)}-v_0^{(k)}{\bf 1}_n\big)^T.
\end{align}

Suppose (\ref{vpq}) holds, i.e.,
\begin{align}
        {\bf G}^{(k)}{\bf y}^\ast=2{\bm\Phi}{\bf p}^{(k)}+2{\bf q}^{(k)},\,\forall k.
\end{align}

Then, (\ref{ek}) is rewritten as,
\begin{align}
\nonumber{\bf e}^{(k)}({\bf y},\lambda)&=\big({\bf v}^{(k)}-v_0^{(k)}{\bf 1}_n\big)-2\lambda{\bf p}^{(k)}-2{\bf q}^{(k)}\\
%\nonumber&={\bf J}^{(k)}{\bf u}-2\lambda{\bf p}^{(k)}-2{\bf q}^{(k)}\\
&={\bf G}^{(k)}({\bf y}-{\bf y}^{\ast})-2(\lambda{\bf I}_n-\bm\Phi){\bf p}^{(k)}
\end{align}
and accordingly,
\begin{align}\label{eul}
    \nonumber\left\|{\bf e}({\bf y},\lambda)\right\|_2^2=\sum_{k=1}^K\Big\|&{\bf G}^{(k)}({\bf y}-{\bf y}^{\ast})\\
    &-2\big[(\lambda-\lambda^\ast){\bf I}_n-\bm\Delta_\lambda\big]{\bf p}^{(k)}\Big\|_2^2.
\end{align}

Problem (\ref{T0}) can be solved by letting $\nabla_{\bf u,\lambda}\left\|{\bf e}({\bf y},\lambda)\right\|_2^2=0$, which yields a closed-form solution as,% (the QP program is supposed to be well-posed),
\begin{align}
\begin{bmatrix}
 \bf y^\star-y^\ast\\
 \lambda^\star-\lambda^\ast
 \end{bmatrix}=&\left(\sum_{k=1}^{K}{\bf H}^{(k)}\right)^{-1}\left(\sum_{k=1}^K{\bf c}^{(k)}\right)
 \end{align}
 with
 \begin{align}
\nonumber {\bf H}^{(k)}&=\begin{bmatrix}
 ({\bf G}^{(k)})^T{\bf G}^{(k)}&-2({\bf G}^{(k)})^T{\bf p}^{(k)}\\
 -2({\bf p}^{(k)})^T{\bf G}^{(k)}&({\bf p}^{(k)})^T{\bf p}^{(k)}
 \end{bmatrix}\\
\nonumber {\bf c}^{(k)}&=2\begin{bmatrix}
 {\bf G}^{(k)}&
 {\bf p}^{(k)}
 \end{bmatrix}^T{\bm\Delta}_\lambda
 {\bf p}^{(k)}.
\end{align}
Therefore, we have
\begin{align}
\nonumber  \left\|\bf y^\star-y^\ast\right\|&\leq\left\|\left(\sum_{k=1}^{K}{\bf H}^{(k)}\right)^{-1}\right\|\left\|\sum_{k=1}^K{\bf c}^{(k)}\right\|\\
\nonumber &\leq\left\|\left(\sum_{k=1}^{K}{\bf H}^{(k)}\right)^{-1}\right\|\left(\sum_{k=1}^{K}\left\|{\bf c}^{(k)}\right\|\right)\\
&\leq\epsilon\left\|{\bm\Delta}_\lambda\right\|
\end{align}
where
\begin{align}
\nonumber \epsilon=  2\left\|\left(\sum_{k=1}^{K}{\bf H}^{(k)}\right)^{-1}\right\|
\left(\sum_{k=1}^K
\left\|\begin{bmatrix}
 {\bf G}^{(k)}&
 {\bf p}^{(k)}
 \end{bmatrix}^T\right\|\left\|{\bf p}^{(k)}\right\|\right).
\end{align}
This concludes the proof.\qed
\ifCLASSOPTIONcaptionsoff
  \newpage
\fi



\bibliographystyle{IEEEtran}
\bibliography{IEEEabrv,./bibtex/bib/IEEEexample}

\end{document}